\documentclass[aps,prd,twocolumn,nobalancelastpage,superscriptaddress,showpacs]{revtex4-1}
\usepackage[english]{babel}
\usepackage{amsmath,amssymb}
\usepackage[usenames]{color}
\usepackage[pdftex]{graphicx}
\newcommand{\bc}{\begin{center}}
\newcommand{\ec}{\end{center}}
\newcommand{\bt}{\begin{tabular}}
\newcommand{\et}{\end{tabular}}
\newcommand{\be}{\begin{equation}}
\newcommand{\ee}{\end{equation}}
\newcommand{\bea}{\begin{eqnarray}}
\newcommand{\eea}{\end{eqnarray}}
\newcommand{\bfig}{\begin{figure}}
\newcommand{\efig}{\end{figure}}
\newcommand{\ie}{{\it i.e. }}
\begin{document}

\title{Is the Dark Disc contribution to Dark Matter Signals important ?}

\author{Fu-Sin Ling}
\affiliation{Service de Physique Th\'eorique,
  Universit\'e Libre de Bruxelles,
  Boulevard du Triomphe - CP225,
  1050 Bruxelles - Belgique}
\email{fling@ulb.ac.be}

\begin{abstract}
Recent N-body simulations indicate that a thick disc of dark matter, co-rotating with the stellar disc, 
forms in a galactic halo after a merger at a redshift $z<2$.
The existence of such a dark disc component in the Milky Way could affect dramatically dark matter 
signals in direct and indirect detection.  
In this letter, we discuss the possible signal enhancement in connection with the characteristics 
of the local velocity distributions.
We argue that the enhancement is rather mild, but some subtle effects may arise.
In particular, the annual modulation observed by DAMA becomes less constrained by other 
direct detection experiments.
\end{abstract}


\maketitle

\begin{flushleft}
Preprint: ULB-TH/09-35
\end{flushleft}
			   		  
\section{Introduction}
\label{sec:intro}

The problem of determining the local phase-space structure of Dark Matter (DM) in the solar 
neighborhood is rather difficult.
The distance of the Sun to the galactic center is $R_0 = 8.0$~kpc, 
which represents only a small fraction (around 3--4\%)
of the virial radius of the Milky Way galaxy.
One would therefore expect most DM particles to have virialized, 
with only a very small fraction in cold flows, \ie flows with 
negligible velocity dispersion~\cite{Sikivie:1995dp,Ling:2004aj}.
This is indeed confirmed by high resolution N-body simulations, 
which find only faint evidence for caustics~\cite{Diemand:2008gf}.

However, the distribution of DM in galactic halos is not smooth. 
Hierarchical clustering in the $\Lambda$CDM paradigm leads to nested structures with numerous subhalos~\cite{Diemand:2009bm}.
Although recent DM only simulations~\cite{Springel:2008cc,Diemand:2008in} have reached a tremendous mass resolution, 
they are still far from resolving the smallest substructures that are expected on theoretical grounds, 
with a mass in the range $10^{-11} \cdots 10^{-3}$ solar masses, depending on the decoupling temperature of the WIMP~\cite{Berezinsky:2003vn,Green:2005fa,Profumo:2006bv,Bringmann:2009vf}.

When baryons enter the game, the presence of an already well formed galactic disc at high redshift perturbs the merger process
of the host galaxy with smaller satellites, as it causes a preferential drag in its direction~\cite{Read:2009iv}. 
As a result, the material from the accreted structure, including DM, stars and gas, settles around the galactic disc of the host halo,
after being disrupted by tidal forces.
In particular, the accreted DM forms a thick disc, often dubbed the \emph{dark disc}, which co-rotates with the stellar disc.

A consequence of this scenario is the formation of a thick disc of stars in the host galaxy, as the result of the heating of the pre-existing
thin disc of stars during the merger, supplemented by the accreted stars.
Many morphological and dynamical properties of the accreted stars and DM should therefore 
evidence some correlation~\cite{Read:2008fh,Purcell:2009yp}.
Using three disc galaxies from cosmological hydrodynamics simulations, that are intended to span a large range of merger histories,
from very quiescent to very violent, Read {\it et al.}~\cite{Read:2009iv} conclude that
the existence of a thick stellar disk and a co-rotating dark disk in the $\Lambda$CDM paradigm is quite robust.

The detailed quantitative properties of the final galaxy depend however on its merger history.
In particular, the authors find that the dark disc can contribute $\sim 0.25 - 1.5$ times the non-rotating halo density at the solar position.
Also, the rotation lag of the rotating component is in the range $0-150$~km/s. 
A large value of the rotation lag, corresponding to a small halo circular speed, is only found
for galaxies which had no significant merger after a redshift $z=2$.

The dramatic influence of the dark disk on DM signals has been shown in Ref.~\cite{Bruch:2008rx} for direct detection and 
in Ref.~\cite{Bruch:2009rp} for indirect detection.
For direct detection, the low velocity of the dark disc with respect to the Earth enhances detection rates at low recoil energy.
This could lead to an increase by up to a factor of 3 in the $5-20$~keV recoil energy range, for DM masses bigger than 50~GeV.
The annual modulation is also affected, with a reduction of the phase corresponding to maximal rate by up to three weeks.
For neutrino indirect detection, the boost in signal is even more impressive. The enhanced capture rate in the Sun and in the Earth
leads to a muon flux from the Sun (from the Earth) that can be one (three) orders of magnitude higher than in the Standard Maxwellian halo case.

To arrive to their conclusions, the authors in Ref.~\cite{Bruch:2008rx} and Ref.~\cite{Bruch:2009rp} assumed a lag velocity $v_{lag} = 50$~km/s,
and supposed that the velocity distribution of the dark disc can be correctly described by an isotropic Maxwellian distribution 
with a velocity dispersion $\sigma = 50 {\rm km/s}$, neglecting  anisotropies and deviations from Maxwell-Boltzmann statistics.
and  also taken to have the same value .
They argue that this value of the velocity dispersion is supported by the observed kinematics of the Milky Way thick disc stars,
$(\sigma_R,\sigma_\phi,\sigma_z) = (63,39,39)\pm 4$~km/s~\cite{Soubiran:2002sf},
and is compatible with the results of numerical simulations.
The purpose of this letter is to discuss whether these assumptions are realistic or not, and what would the implications be for
direct and indirect detection.

\section{Dynamical properties of the dark disc}
\label{sec:DD}    

\begin{figure*}[t]
\begin{center}
\begin{tabular}{ccc}
\includegraphics[width=0.97\columnwidth, clip]{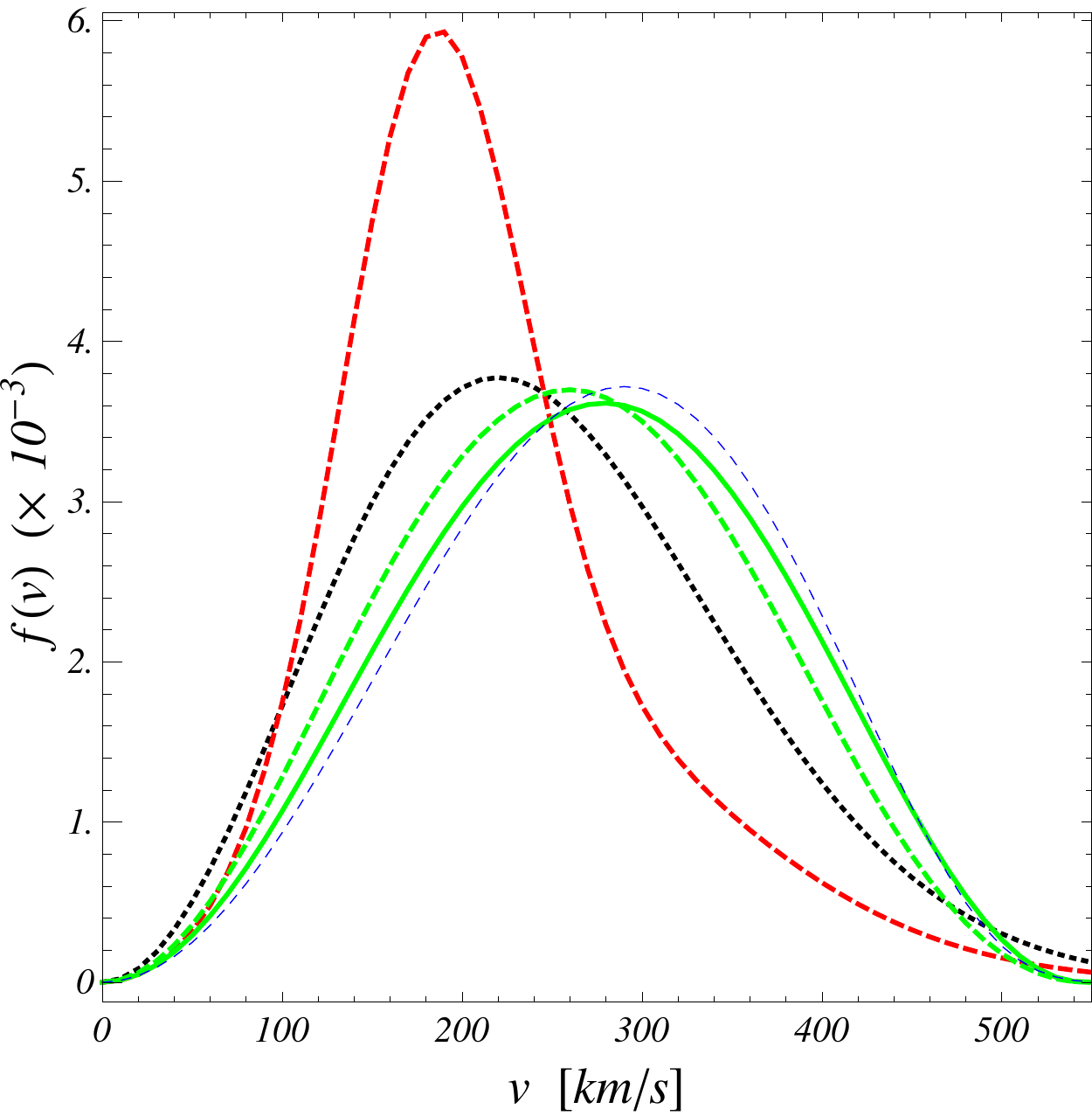}& ~~ &
\includegraphics[width=\columnwidth, clip]{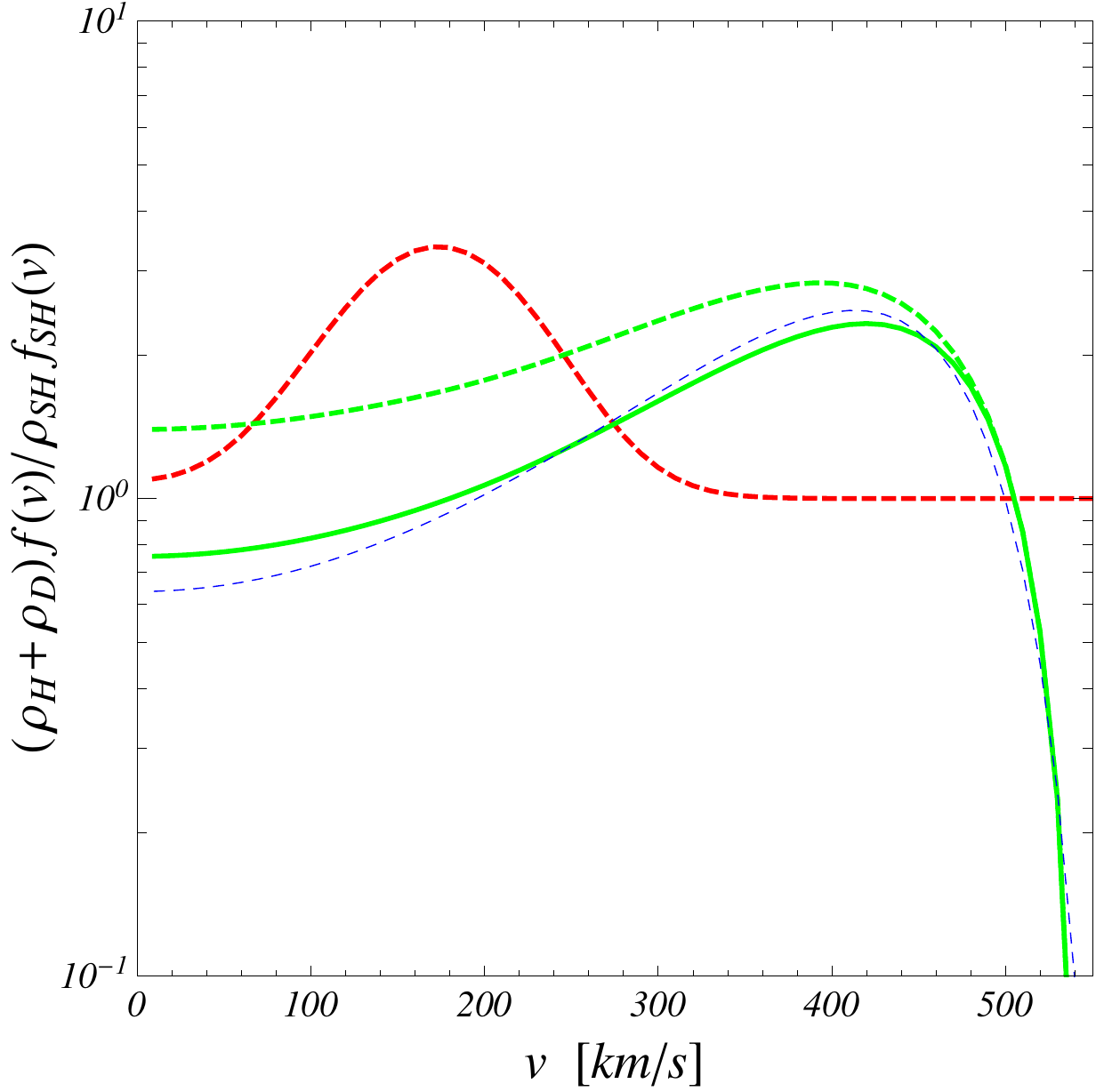}\\
\includegraphics[width=0.97\columnwidth, clip]{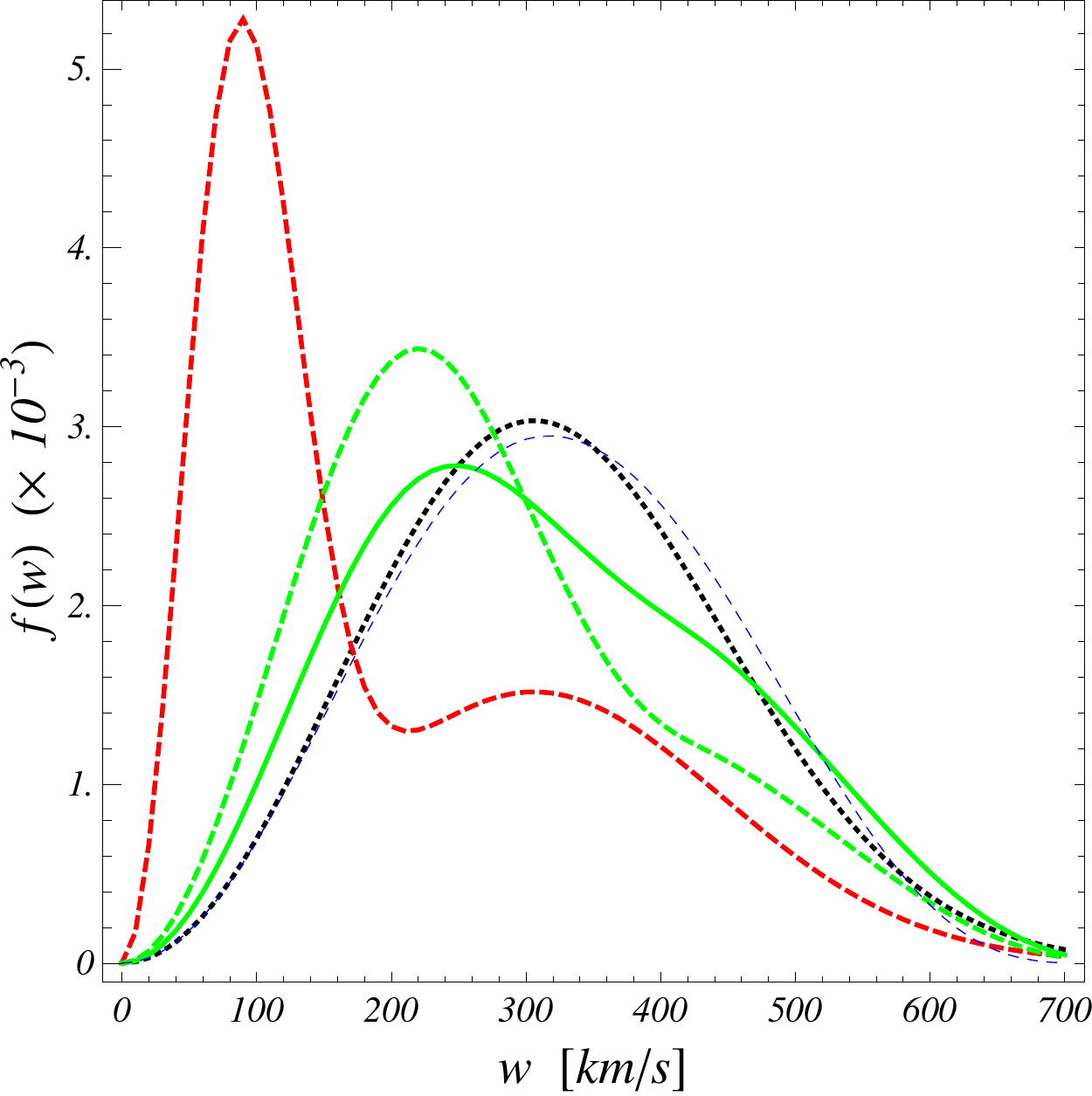}& ~~ &
\includegraphics[width=\columnwidth, clip]{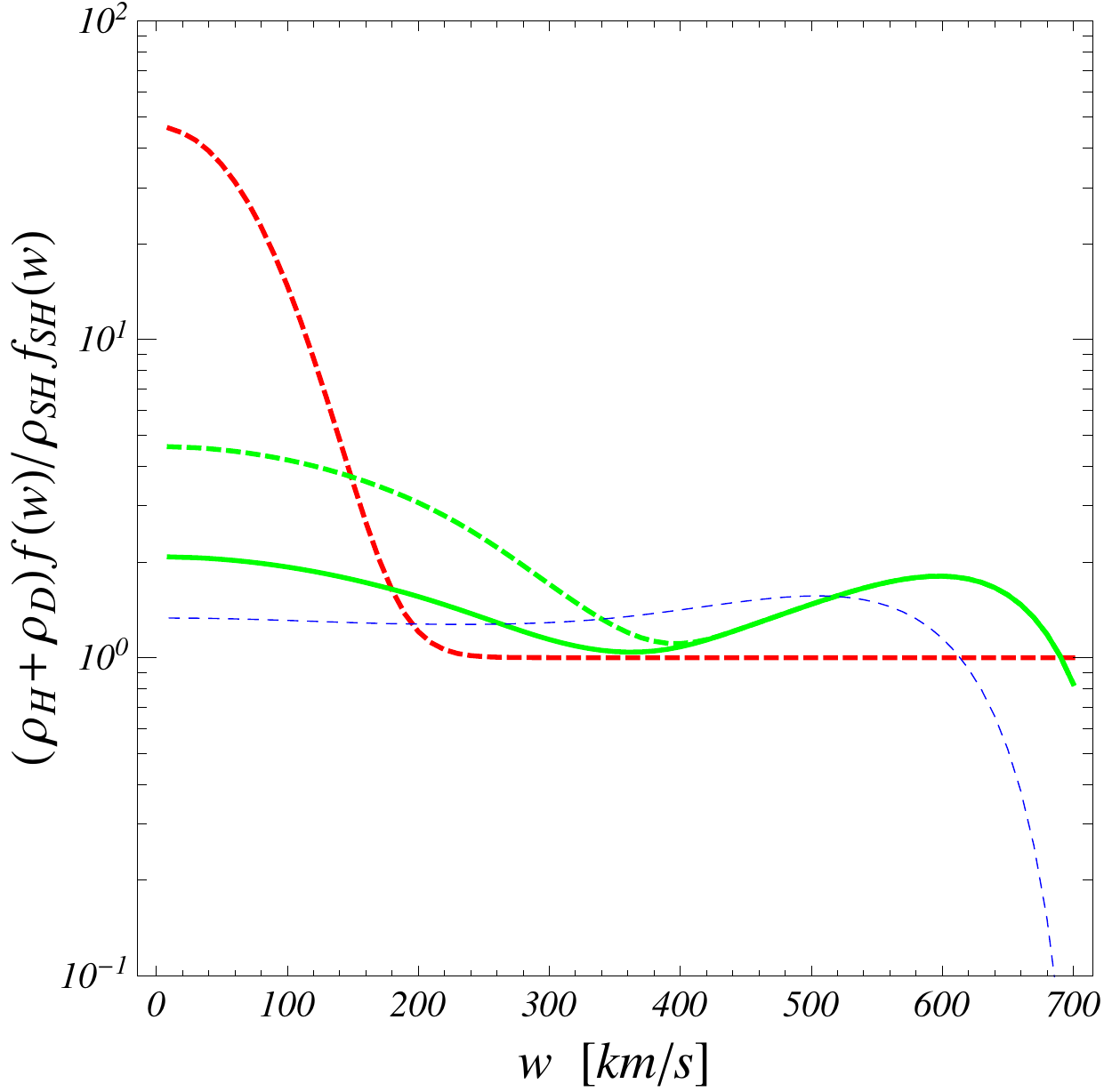}\\
\end{tabular}
\caption{\small \it 
(Left column) Normalized velocity distributions for different dark matter halos, in the galactic frame (top) and 
with respect to the Sun (bottom). 
\emph{Black dotted} : standard Maxwellian halo, 
\emph{Blue thin dashed} : slowly rotating Tsallis halo,
\emph{Red dashed} : halo with a strong dark disc,
\emph{Green solid} : halo with a mild dark disc and $\rho_D/\rho_H = 1/3$,
\emph{Green dashed} : halo with a mild dark disc and $\rho_D/\rho_H = 1/1$.
(Right column) Enhancement of the phase-space distribution compared to the standard Maxwellian halo, for the same halos, 
as a function of the velocity. See text (Section III) for the parameters of the Tsallis, strong and mild dark disc halo models.
}
\label{fig:fw}
\end{center}
\end{figure*}

The dark disc affects the predictions of direct and indirect detection signals essentially because it increases the
number of dark matter particles that move slowly with respect to the Earth or the Sun.
The most relevant parameters that determine the phenomenological importance of the dark disc are therefore
its local density $\rho_D$, its lag velocity $v_{lag}$ with respect to the stellar disc, and
its velocity dispersion $\sigma$. 

One can estimate the velocity dispersion by applying the virial theorem. For a galaxy of the size of the Milky-Way,
\be
\langle v^2 \rangle = \frac{|W|}{M} \simeq (200~{\rm km/s})^2 \quad , 
\ee
where $W$ is the gravitational potential energy and $M$ is the mass of the galaxy~\cite{Binney:1988}.
Therefore, if we assume a universal density profile for DM halos, a satellite halo with a mass $M_s$ around ten times smaller than the host halo mass $M_h$
should have a velocity dispersion given by
\be
\langle v^2 \rangle_s \simeq \left( \frac{M_s}{M_h} \right)^{2/3} (200~{\rm km/s})^2 \quad ,
\ee
which gives an average one dimensional velocity dispersion around 50~km/s.
This value is the mean velocity dispersion for the entire satellite halo, the question is to know how it is related to the velocity dispersion 
of the accreted DM at the Earth's location after merging of the satellite halo.
The merging and disruption processes are likely to cause some heating. Also, the final velocity dispersion will depend
on whether the accreted DM originates rather from the inner or the outer parts of the satellite halo,
as the average velocity of the particles drop  with the distance from the center of the halo.

In Ref.~\cite{Read:2009iv}, three cosmological hydrodynamics simulations of bright, disc dominated galaxies are used to
predict the phase-space distribution of dark matter in the solar neighborhood.
The three galaxies have been chosen to span very different merger histories, from a very quiescent one after a redshift $z=2$ to 
a violent one with a massive merger after redshift $z=1$. 
This sample is meant to cover the possible extreme cases of the merger history of the Milky Way, and to bracket the 
characteristics of its dark disk.
Without surprise, the contribution of the dark disc to the local dark matter density near the Sun increases
when there are more accretion events. The authors find that $\rho_D = 0.25 \dots 1.5 \rho_H$, where $\rho_H$ is the local density
of the host halo. The lag velocity, between 0~and 150~km/s, is also smaller for more massive mergers.

Interestingly, on Fig.~4 of Ref.~\cite{Read:2009iv}, the authors compare the distributions of \emph{accreted} stars and dark matter.
It appears that the accreted DM velocity dispersion is larger than 100~km/s for all three galaxies considered in their paper.
More precisely, the quadratic average of the velocity dispersion in the directions $R,\phi,z$ amounts to
$\sigma = 119.8$, $\sigma = 108.1$ and $\sigma = 135.7$~km/s for the galaxies MW1, H204 and H258 respectively.
For accreted stars, the average velocity dispersion is much smaller, with the largest value in the radial direction, 
$\sigma = 85.3$, $\sigma = 83.5$ and $\sigma = 90.7$~km/s for MW1, H204 and H258 respectively.
As we can see, the velocity dispersion of the accreted DM is at least 25\% higher than that of the accreted stars.
This result is in agreement with the conclusions of Ref.~\cite{Purcell:2009yp}, where the authors use
focused, high resolution simulations of accretion events to compare the morphological and kinematical properties 
of the dark disc with those of the Milky Way's thick disc of stars.

In Ref.~\cite{Purcell:2009yp}, single accretion events at different orbital infall inclinations
are studied with focused numerical simulations.
It appears that the velocity dispersion of accreted stars is substantially larger than 
the velocity dispersion of the stars in the resultant thick disc of the host halo,
except for very low inclination mergers.
However, these planar events produce stellar discs that are much hotter than what is observed for the 
Milky Way~\cite{Purcell:2009yp}. Indeed, the thick disc of the Milky Way has a velocity dispersion 
$(\sigma_R,\sigma_\phi,\sigma_z) = (63 \pm 6, 39 \pm 4, 39 \pm 4)$~km/s, it also has a 
moderate rotational lag $v_{lag}=51 \pm 5$~km/s~\cite{Soubiran:2002sf}.
For higher inclination accretion events ($\theta \sim 60^\circ$), the resultant thick disc is less hot and compatible
with observations. However, in these cases, the accreted stars have a velocity dispersion $2-3$ times larger than that 
of the thick disc, which results from the heating of the primary disc.
Therefore, taking the observed values of the velocity dispersions of the Milky Way thick disc as a way to characterize
the dark disc velocity dispersion is unlikely to be realistic. 
Nevertheless, this provides a very conservative lower bound 
$1.25 (\sigma_R,\sigma_\phi,\sigma_z) \simeq (89, 50, 50)$~km/s.

More realistic values of the dark disc velocity dispersion $\sigma_D$ are provided by full cosmological hydrodynamics simulations.
We already mentioned that $\sigma_D$ exceeds 100~km/s for all three galaxies considered in Ref.~\cite{Read:2009iv}.
Another high resolution simulation with baryons has been analyzed in Ref.~\cite{Ling:2009eh}.
There, the velocity distribution along $\phi$ for DM particles in a ring $7<R<9$~kpc, $|z|<1$~kpc showed that the fraction 
of particles co-rotating with the stellar disk with a small lag velocity cannot be large.
A fit with a double Gaussian distribution gave the following possible parameters for the dark disc component:
$\rho_D = 0.25 \rho_H$, $v_{lag} = 70$~km/s, and $\sigma_\phi = 85$~km/s.
At this point, it is important to stress that the double Gaussian fit gives an acceptable description of the simulation 
velocity distribution only in the direction $\phi$.
Indeed, if one infers isotropic velocity dispersions $\sigma_D$ for the dark disc and $\sigma_H$ for the halo from this fit,
such extrapolation would lead to velocity distributions in the radial and $z$ directions that are much more peaked
than the actual ones. They would be leptokurtic, instead of platykurtic, as seen in the simulation~\cite{Ling:2009eh}.

The failure of this extraction and extrapolation is related to the fact that equilibrated dark matter haloes show systematic deviations 
from Maxwell-Boltzmann statistics. The presence of a long-range gravitational force suggests that this kind of system is better
described by the nonextensive Tsallis statistics~\cite{tsallis-88,Hansen:2004dg}. 
Indeed, for the DM particles in a spherical shell around 8~kpc in Ref.~\cite{Ling:2009eh}, 
the velocity distribution is very well fit by a Tsallis distribution with parameters $q \simeq 0.8$
and $v_0=265$~km/s. We recall that the Tsallis distribution is defined as
\be
{\rm f}(\vec{v}) = \frac{1}{N(v_0,q)} \left(1-(1-q) \frac{\vec{v}^2}{v_0^2} \right)^{q/(1-q)} \quad ,
\ee
where $N(v_0,q)$ is a normalization factor.
For ring particles with $7<R<9$~kpc, $|z|<1$~kpc, we can get a reasonable fit of the velocity distributions as a sum of two isotropic 
Tsallis distributions if we take $q_H = q_D = 0.7$, $v_0^H=300$~km/s, $v_0^D=200$~km/s, and $v_{lag}=70$~km/s.
With this decomposition, the dark disc velocity dispersion is $\sigma_D = 117$~km/s, much larger than that inferred from the double Gaussian fit.
The same discrepancy also appears in Ref.~\cite{Read:2009iv}, as a double Gaussian fit of the $v_\phi$ distributions (see Fig. 2 of Ref.~\cite{Read:2009iv})
gives $\sigma_D = 50.3$, $\sigma_D = 76.0$, and $\sigma_D = 87.9$~km/s for galaxies MW1, H204, and H258 respectively,
values that are more than 30\% smaller than the average velocity dispersions of the accreted dark matter (see above).
In other words, the double Gaussian fit of the $v_\phi$ distribution underestimates the  dark disc velocity dispersion.

We conclude that the velocity dispersion of the Milky Way's dark disc is likely to be larger than 100~km/s. As we have shown, this is indeed the case 
for all the galactic halos comparable to the Milky Way that are extracted from full cosmological simulations with baryons. 
These halos have undergone several accretion events during their merger history, which might explain why the resultant dark disc velocity distribution
is hotter than in some focused and controlled single accretion event simulations.

\section{Dark matter signals from the Dark Disc}
\label{sec:signals}

\begin{figure*}[t]
\begin{center}
\begin{tabular}{ccc}
\includegraphics[width=\columnwidth, clip]{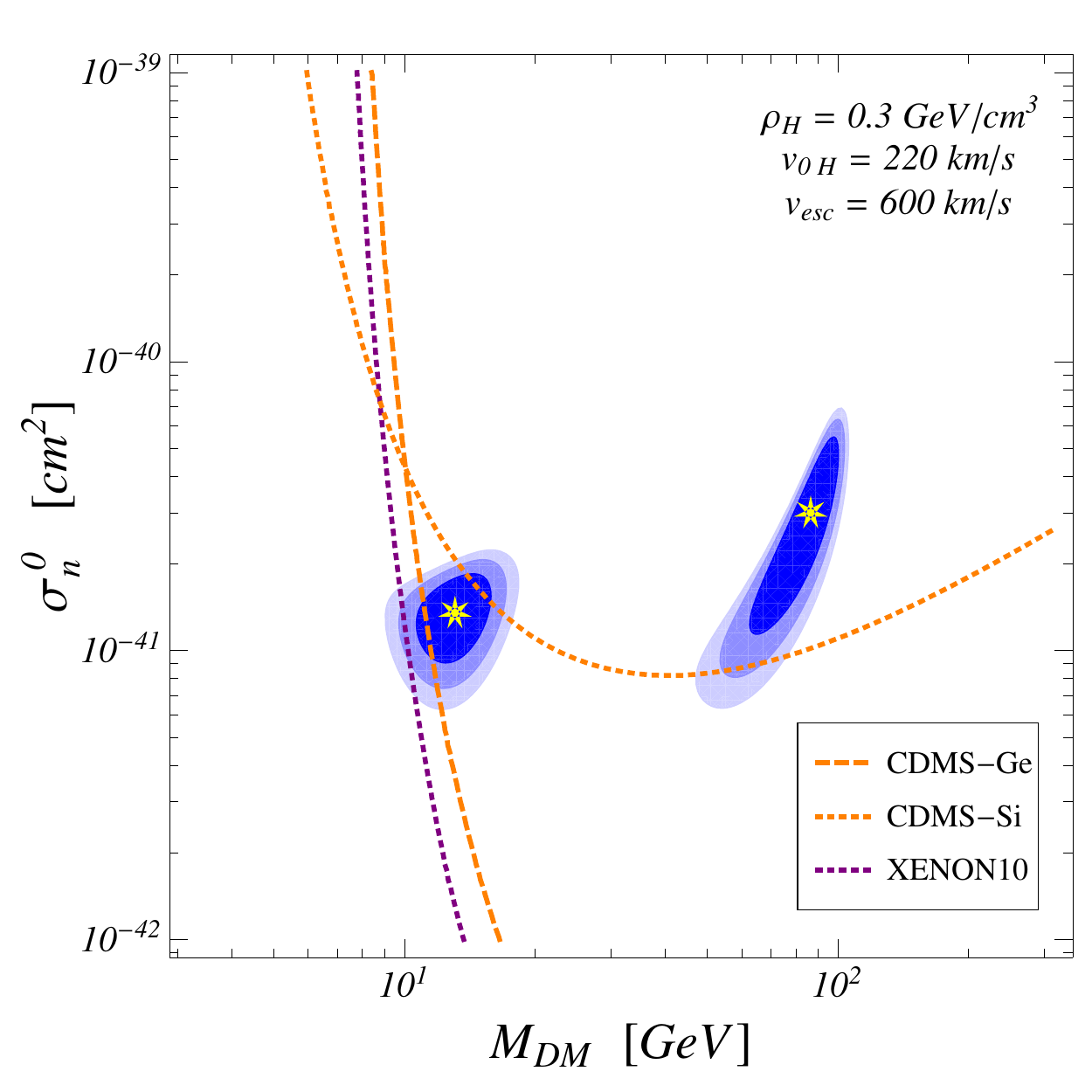}& ~~ &
\includegraphics[width=\columnwidth, clip]{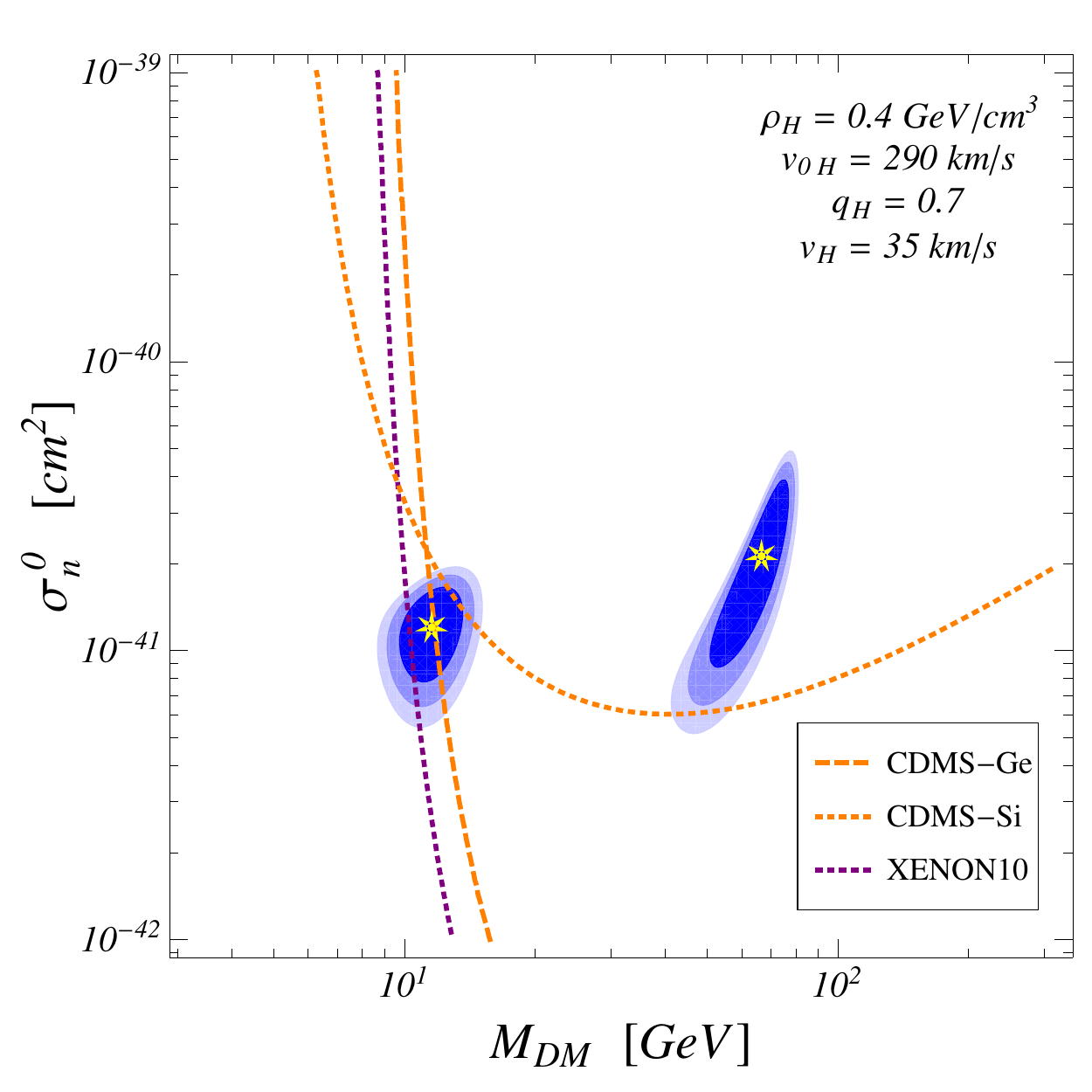}\\
\includegraphics[width=\columnwidth, clip]{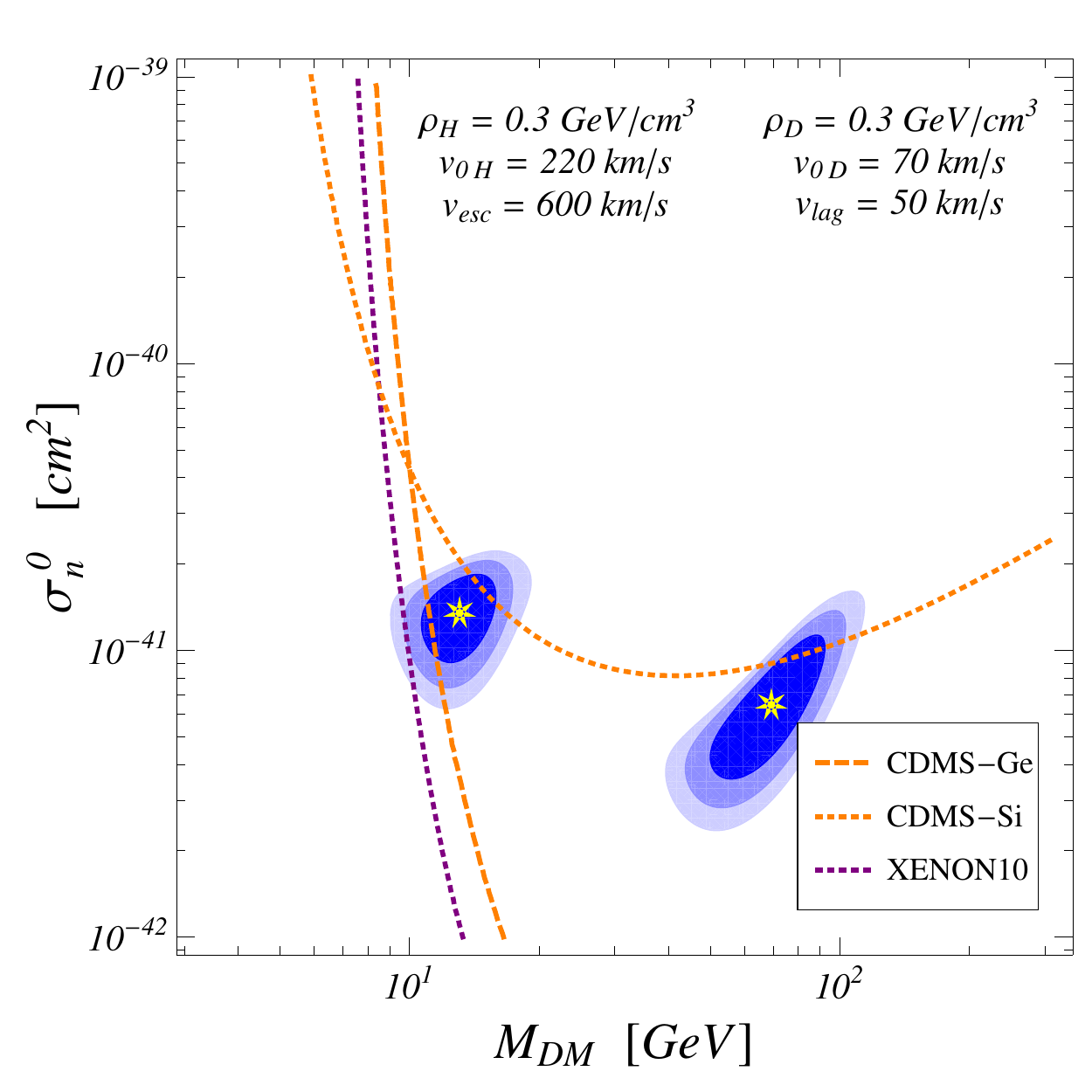}& ~~ &
\includegraphics[width=\columnwidth, clip]{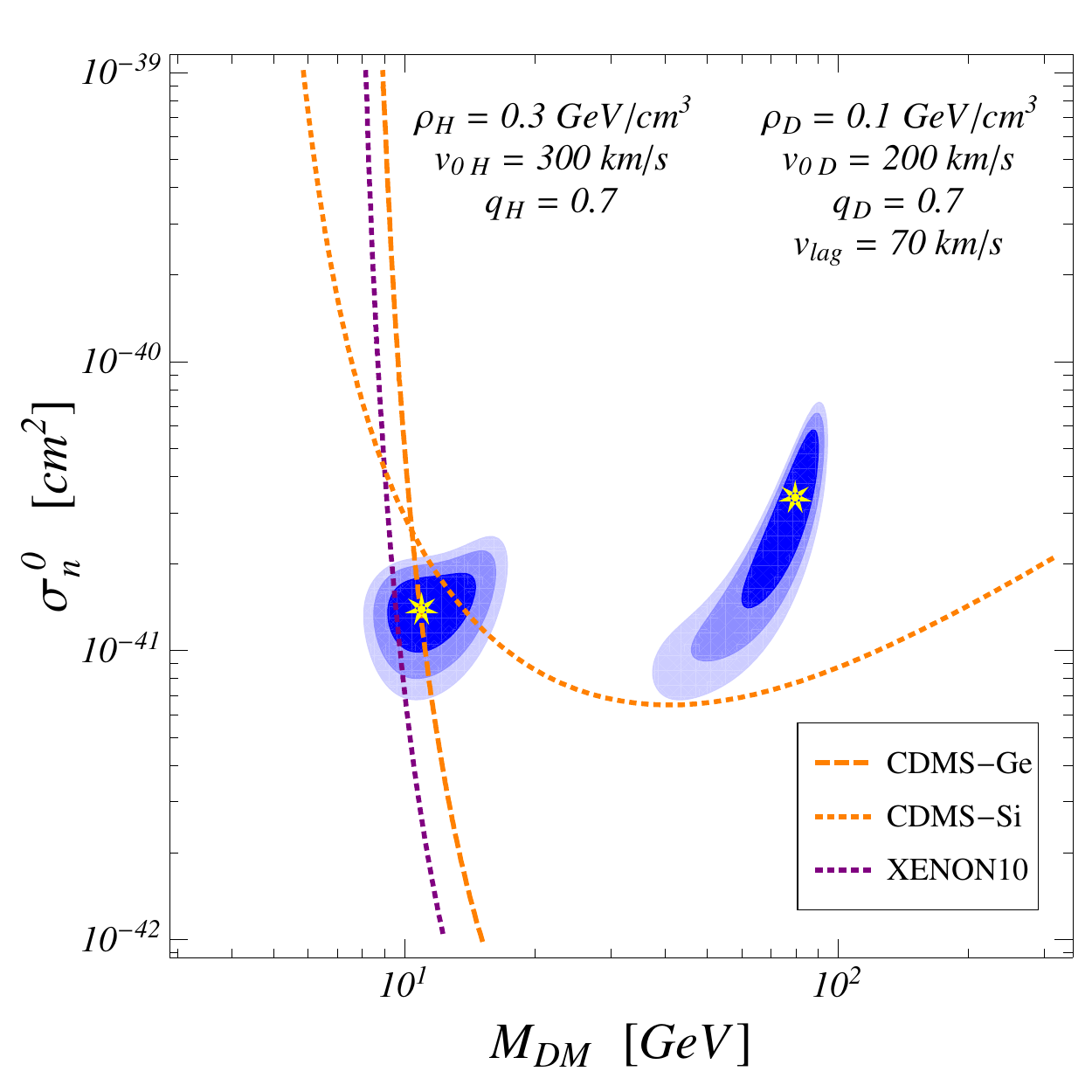}\\
\end{tabular}
\caption{\small \it 
{\rm \underline{Direct detection - Elastic scenario}} :  
DAMA allowed regions in the elastic scenario compared with exclusion limits of CDMS and XENON, in the case of
a standard Maxwellian halo (top left), a slowly co-rotating Tsallis halo (top right),
a strong Maxwellian dark disc halo (bottom left), and a mild Tsallis dark disc halo (bottom right).
DAMA contours are given at the $90$, $99$ and $99.9$\% CL. Stars indicate local best-fit points.
All other exclusion curves are at the $99.9$\% CL.
}
\label{fig:DAMA}
\end{center}
\end{figure*}
\begin{figure*}[t]
\begin{center}
\begin{tabular}{ccc}
\includegraphics[width=\columnwidth, clip]{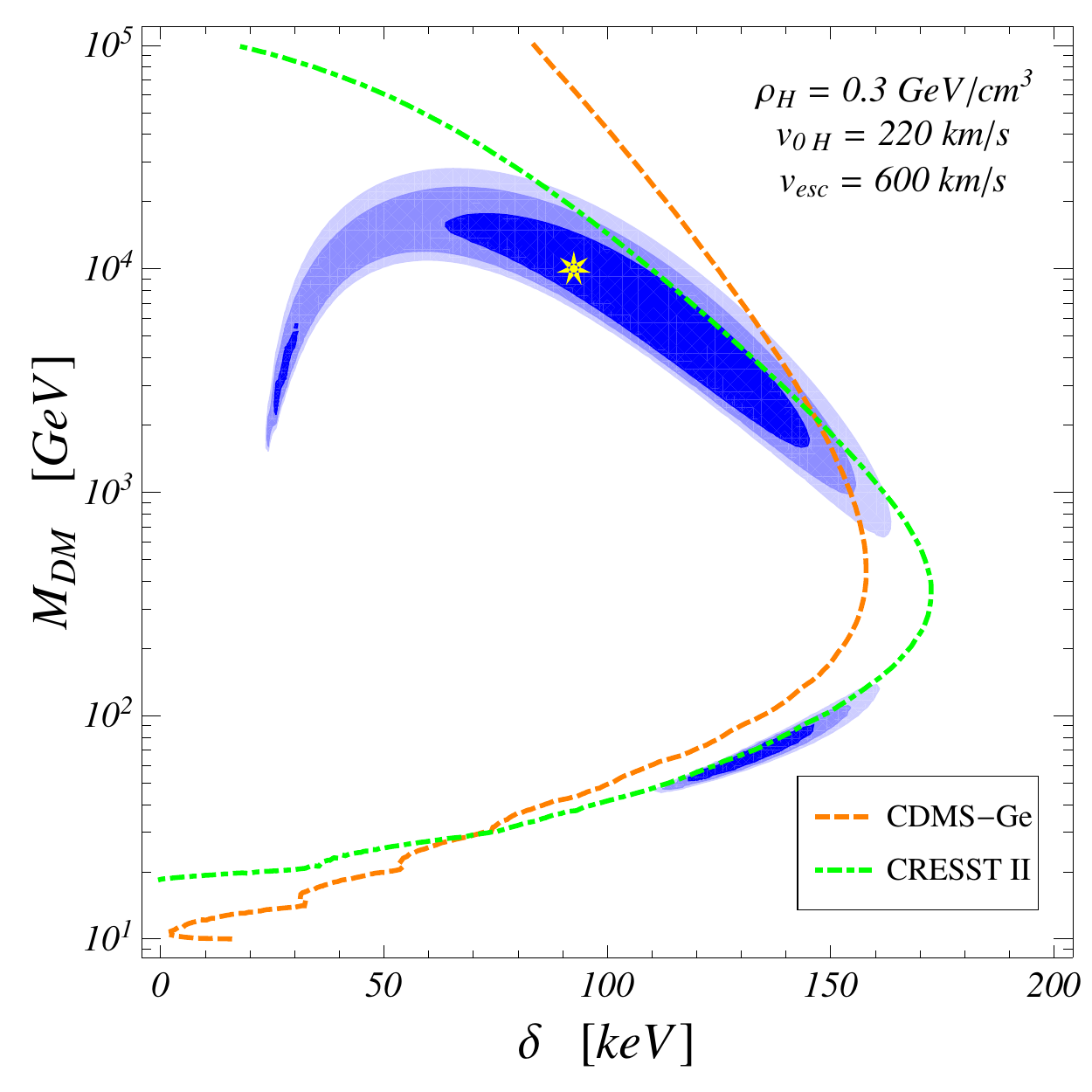}& ~~ &
\includegraphics[width=\columnwidth, clip]{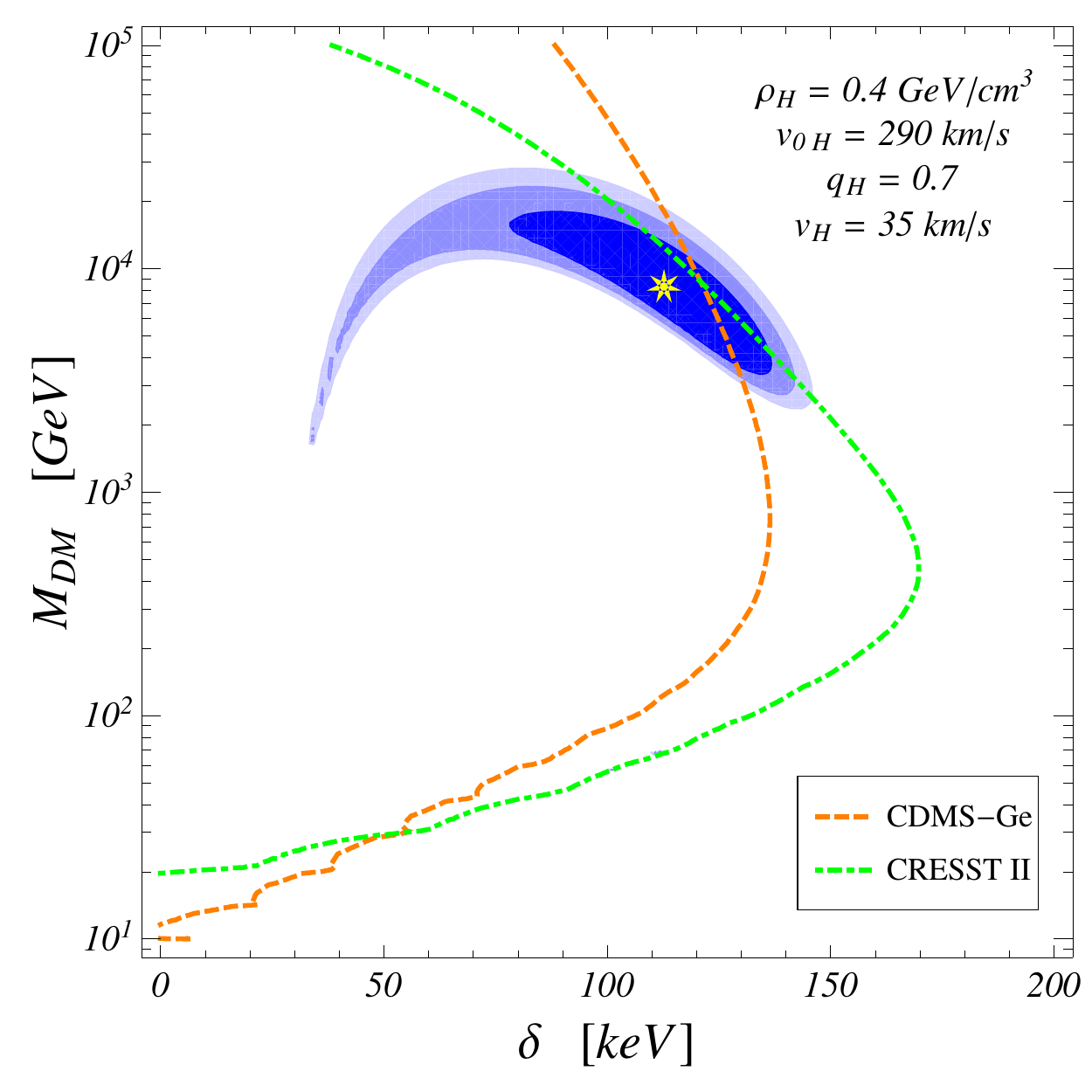}\\
\includegraphics[width=\columnwidth, clip]{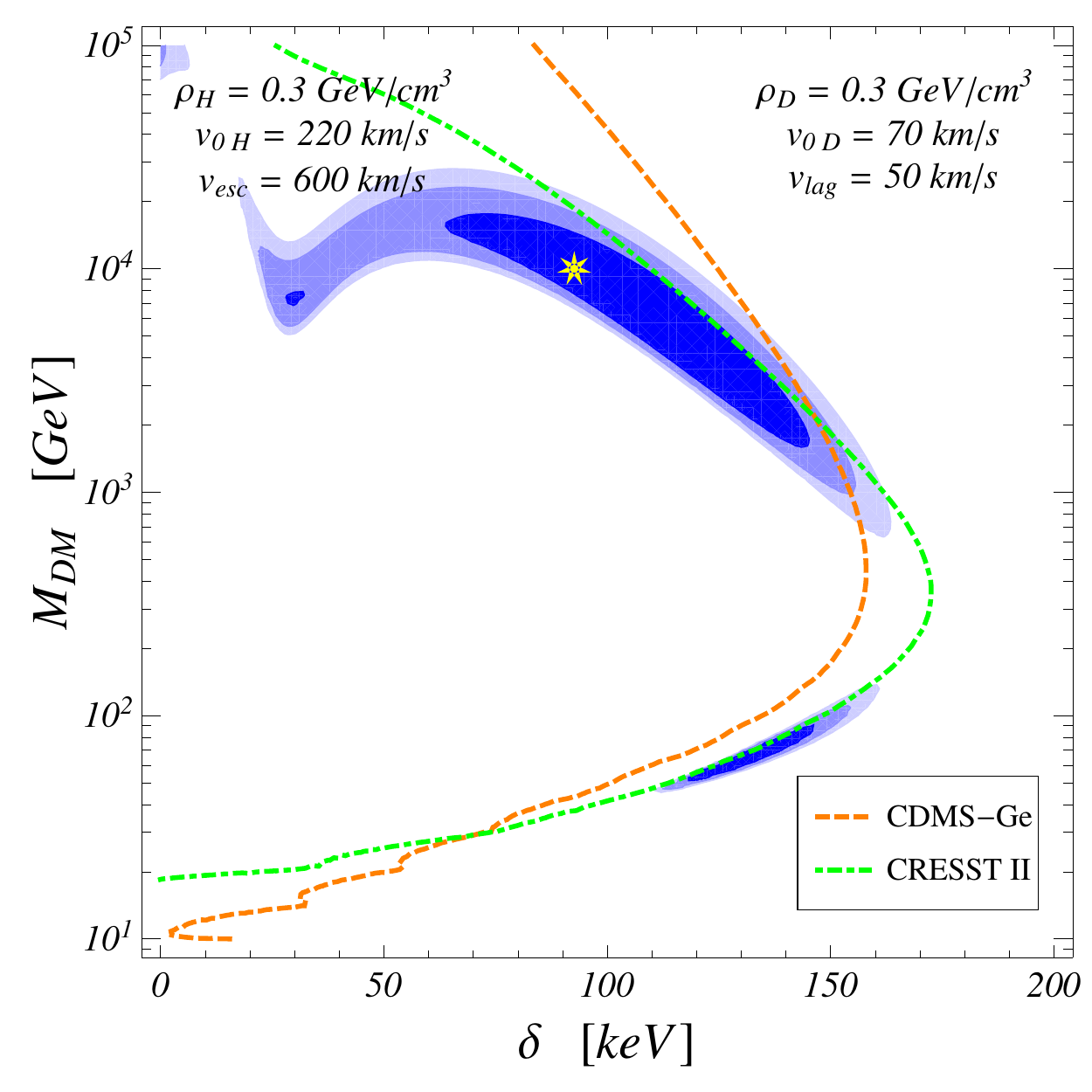}& ~~ &
\includegraphics[width=\columnwidth, clip]{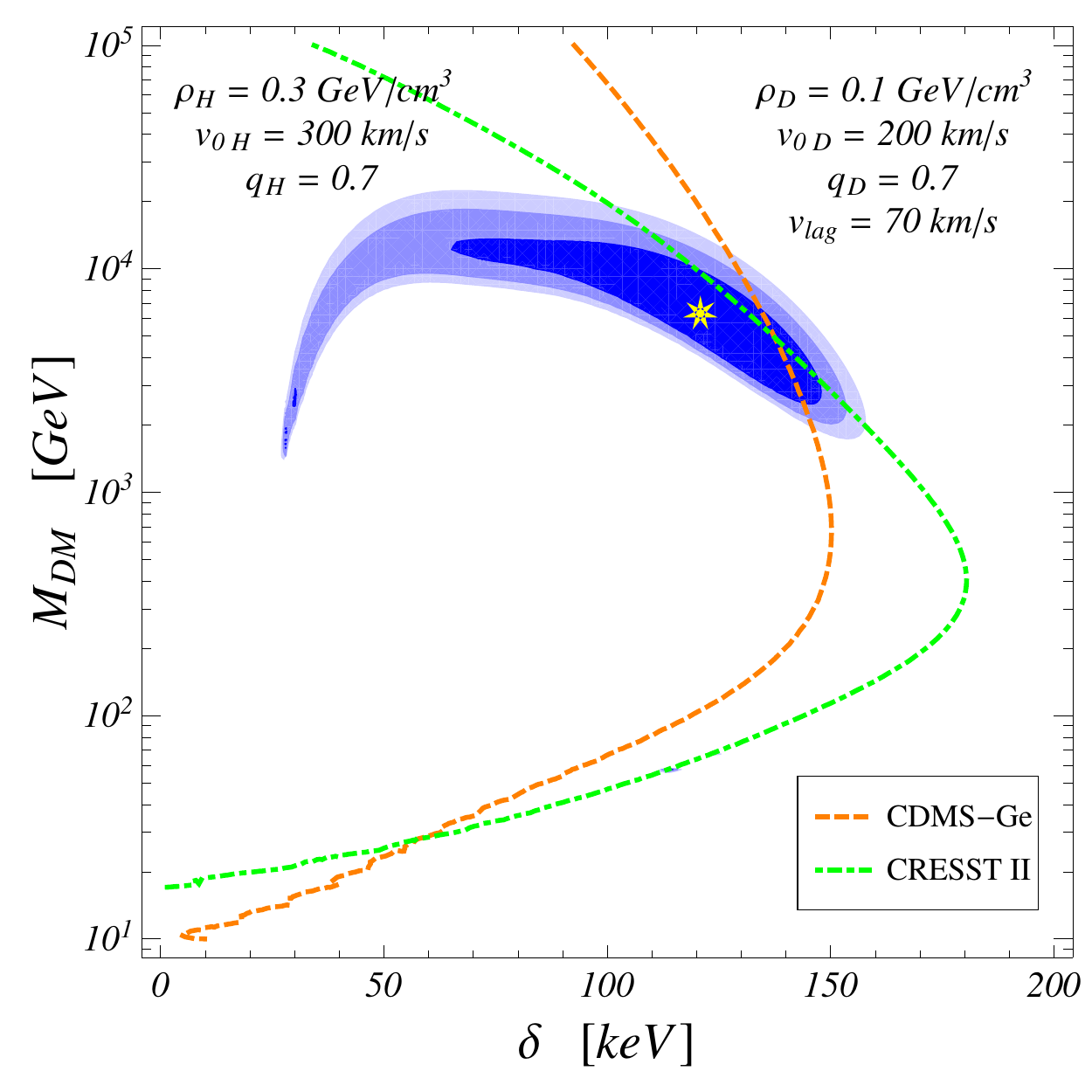}\\
\end{tabular}
\caption{\small \it 
{\rm \underline{Direct detection - Inelastic scenario}} :  
DAMA allowed regions in the inelastic scenario compared with exclusion limits of CDMS and XENON, in the case of
a standard Maxwellian halo (top left), a slowly co-rotating Tsallis halo (top right),
a strong Maxwellian dark disc halo (bottom left), and a mild Tsallis dark disc halo (bottom right).
DAMA contours are given at the $90$, $99$ and $99.9$\% CL. Stars indicate local best-fit points.
All other exclusion curves are at the $99.9$\% CL.
}
\label{fig:DAMA2}
\end{center}
\end{figure*}

The presence of a dark disc in the galaxy can lead to an enhancement of dark matter signals in both direct and indirect detection.
Much of this enhancement stems from an excess, compared to the standard Maxwellian halo predictions, of particles
with a small relative velocity with respect to the Sun (and the Earth).  
The contribution of the dark disc to dark matter signals has been calculated in Ref.~\cite{Bruch:2008rx} for direct detection,
and in Ref.~\cite{Bruch:2009rp} for indirect detection.
Both papers assumed a Maxwellian profile (isotropic) for both the static halo and the rotating dark disc, with parameters
$(v_0^H,v_0^D,v_{lag})=(220,50 \sqrt{2},50)~{\rm km/s}$.
As discussed in Sec.~\ref{sec:DD}, the Milky Way's dark disc is unlikely to have such a small velocity dispersion.
In the sequel, we will refer to this parameterization as the \emph{strong} dark disc model, for which we also assume
$\rho_D/\rho_H = 1$ and $\rho_H = 0.3~{\rm GeV/cm^3}$. So, in this model, the host halo is the standard Maxwellian halo. 

We also consider a \emph{mild} dark disc situation, which models well the velocity distributions extracted from Ref.~\cite{Ling:2009eh}.
We take isotropic Tsallis distributions for both the static halo and the rotating dark disc, with parameters $q_H = q_D = 0.7$ and
$(v_0^H,v_0^D,v_{lag})=(300,200,70)~{\rm km/s}$. Here, we let the dark disc to halo density ratio have two values, $\rho_D/\rho_H = 1/3$
as in the simulation, or $\rho_D/\rho_H = 1$, in order to compare with the strong dark disc model.
For the mild dark disk case, the velocity dispersion is much larger $\sigma_D = 117$~km/s. We recall that such a large value
is supported by hydrodynamics cosmological simulations~\cite{Purcell:2009yp,Read:2009iv} and 
kinematical properties of the Milky Way's thick disc~\cite{Soubiran:2002sf}.

Finally, in order to separate the effects due to the adoption of Tsallis distributions from those of the dark disc, 
we also consider a Tsallis halo with parameters $v_0^H = 290~{\rm km/s}$, $q_H = 0.7$, $\rho_H = 0.4~{\rm GeV/cm^3}$,
which is globally co-rotating with the stellar disc with a speed $v_H = 35~{\rm km/s}$, so that this model halo
has the same average co-rotation velocity as the mild dark disc model halo with $\rho_D/\rho_H = 1/3$.

The velocity distributions of all these models are shown on Fig.~\ref{fig:fw}, in both the galactic frame
and the Sun-based frame, with $\vec{w}=\vec{v}-\vec{v}_\odot$ and $\vec{v}_\odot=(10,225,7)$~km/s 
is the velocity of the Sun in galactic coordinates~\cite{Dehnen:1997cq}. 
The rotating dark disc leads to an increase of particles moving slowly with respect to the Sun.
As we can see, the enhancement is quite sensitive to the dark disc characteristics. 
While for the strong dark disc, the enhancement factor in the limit $w \rightarrow 0$ is about 50,
for the mild dark disc, it hardly reaches 4.5 when the two dark discs have the same local density.
With the realistic ratio $\rho_D/\rho_H = 1/3$, the enhancement factor is about 2.
For the mild dark disc models, there is also a significant bump at higher velocities, 
it is due to the shape of the Tsallis distributions with $q<1$ which are flatter and fatter 
than the Maxwellian distributions. 
It is worth noticing that the bump is due to the static halo part, as it is not modified
when the density of the dark disc is varied. 
As a result, a Tsallis halo with the same global rotation whose velocity distribution in the galactic frame 
follows closely that of the mild dark disc halo model fails to produce a big bump at high velocity in the Sun-based
distributions (see the respective positions of the blue thin dashed and green solid curves in top and bottom right
panels of Fig.~\ref{fig:fw}). 
However, a static Tsallis halo can nevertheless produce such a high velocity bump as seen in the mild dark disc halo 
model, although the absence of co-rotation of the dark matter halo is less realistic in the presence of a baryonic disc.

\subsection{Direct Detection}

Direct detection experiments are aimed at measuring the nuclear recoil of a WIMP-like DM particle with a fixed target.
The differential event rate of nuclear recoils as a function of the recoil energy $E_R$ is conveniently factored as~\cite{Jungman:1995df}
\begin{equation}
\label{eq:diffrate}
\frac{d \mathcal{R}}{d E_{R}} = \frac{\rho_{DM}}{M_{DM}} \frac{d\sigma}{d E_{R}} \; \eta (E_R,t) \quad ,
\end{equation}
where $\rho_{DM}$ is the local density of DM, $M_{DM}$ is the WIMP mass, $d \sigma/d E_{R}$ encodes all the particle and nuclear physics factors, and
$\eta (E_R,t)$ is the mean inverse velocity of incoming DM particles that can deposit a recoil energy $E_R$. 
The time dependence of the velocity distribution is induced by the motion of the Earth around the Sun, which leads to a 
seasonal modulation of the event rate~\cite{Drukier:1986tm,Freese:1987wu}. 

Here we consider both the elastic and the inelastic scattering scenarios.
The cross-section factor is written as
\be
\frac{d \sigma}{d E_{R}} = \frac{M_N \sigma^0_n}{2 \mu^2_n}\ 
\frac{\Big(Z f_p + (A-Z) f_n \Big)^2}{f_n^2} F^2(E_R) \quad ,
\ee
where $M_N \simeq A M_n$ is the mass of the target nucleus with atomic number $A$ and $Z$ protons, 
$\mu_n$ is the reduced WIMP/nucleon mass, $\sigma^0_n$ is the zero momentum WIMP-neutron effective cross-section, 
$f_p$ and $f_n$ are the effective coherent couplings of the WIMP to the proton and the neutron,
and $F^2(E_R)$ is the nuclear form factor which characterizes the loss of coherence for non zero momentum transfer.
We use for the form factor the simple parameterization given by Helm~\cite{Helm:1956zz,Lewin:1995rx}.
For elastic scattering, we consider scalar interactions for which $f_p \simeq f_n$.
For inelastic scattering, we take the weak interaction $Z$ boson exchange as a typical 
realization~\cite{TuckerSmith:2001hy, Cui:2009xq, Arina:2009um}, with
$f_p/f_n \simeq -0.08$ and $\sigma^0_n = G_F^2 \mu_n^2/2\pi$.

The dependence upon the DM velocity distribution is encoded in the quantity
\begin{equation}
\eta (E_R,t) =  \int_{v_{min}} d^3 \vec{v}\  \frac{ {\rm f}(\vec{{v}})}{{|\vec{v}-\vec{v}_\oplus(t)|}}  \quad ,
\end{equation}
where $\vec{v}$ and $\vec{v}_\oplus(t)$ are the WIMP velocity and the Earth velocity in the galactic frame.
A minimum impact velocity $v_{min}$ is needed to provoke a recoil inside the detector,
\be
v_{min} = \sqrt{\frac{1}{2 M_N E_R}} \Big(\frac{M_N E_R}{\mu}+\delta\Big) \quad ,
\label{eq:vmin}
\ee
with $\mu$ the WIMP-nucleus reduced mass, and $\delta$ is the mass splitting between the dark matter
candidate and the next-to-lightest dark state in the inelastic scenario. 
Eq.~(\ref{eq:vmin}) shows that, in the case of elastic scattering, 
the dark disc contributes to low recoil energy events. It dominates
over the halo signal below some value of the recoil energy which increases with the WIMP mass 
for a fixed target. As this analysis has been carried out in details in Ref.~\cite{Bruch:2008rx} for a
strong dark disc, we will not repeat it here.
In the case of inelastic scattering, the enhancement of the phase-space at low velocities produced
by the dark disc will contribute to the direct detection signal only for very small values
of $\delta$, typically $\delta < 30$~keV. The most interesting range of values is however higher,
$50 < \delta < 150$~keV, as the inelastic scattering scenario was first devised as a new way
to accommodate the annual modulation signal observed by DAMA~\cite{Bernabei:2000qi,Bernabei:2008yi}
with exclusion limits set by other experiments that are compatible with null results.

Fig.~\ref{fig:DAMA} (elastic) and Fig.~\ref{fig:DAMA2} (inelastic) show the DAMA allowed 
regions compared with limits from CDMS-Ge~\cite{Akerib:2004fq,Akerib:2005kh,Ahmed:2008eu},
CDMS-Si~\cite{Akerib:2003px,Akerib:2005kh}, XENON10~\cite{Angle:2007uj,Angle:2008we}
and CRESST II~\cite{Angloher:2004tr,Angloher:2008jj}. 
We refer the reader to these experimental papers as well as to Refs.~\cite{Ling:2009eh,Arina:2009um} 
for a description of each experiment, and for values of the exposures, energy thresholds,
efficiencies, etc. that were used. We just mention that a more conservative limit for XENON10 has been chosen,
following the analysis of Ref.~\cite{Fairbairn:2008gz}.

\subsubsection{Elastic scenario}

In the case of a strong dark disc, the so-called channeling region for DAMA~\cite{Bernabei:2007hw,Bottino:2007qg}, 
with $M_{DM} \sim 10 \dots 20$~GeV, is the same as with the standard Maxwellian halo. 
Indeed, channeled events on Iodine ($A_I = 127$) dominate in this 
region~\cite{Bottino:2007qg,Bottino:2008mf,Petriello:2008jj,Savage:2008er,Fairbairn:2008gz,Chang:2008xa},
so that $v_{min} \sim 400$~km/s in the recoil range $2<E_R<6$~keV where the modulation is observed.
Therefore, the compatibility of DAMA and other exclusion limits is not improved by the dark disc.
The dark disc starts to dominate for DM masses above $\sim 30$ GeV.
Compared to the standard halo case, the so-called quenched region is displaced towards smaller cross-sections.
It is however still totally excluded by other null experiments limits.

In the case of a mild dark disc, the quenched region is only slightly affected.
On the contrary, the flatter profiles of Tsallis distributions affects the channeling region significantly.
This region now clearly protrudes outside the XENON10 exclusion limit towards smaller DM masses.
While the best-fit point of the region is still excluded, some part of the 90\% confidence level (CL) region is not.
The level of improvement in compatibility between DAMA and the other experiments observed for the mild dark disc 
setup and in the case of a globally slowly rotating Tsallis halo is the same, although the exact location of the
channeling region is slightly different. We find that the improvement can be related to how fast the high-velocity
tail of the distribution drops compared to the Maxwellian halo.
So, the channelling region of DAMA can be more than marginally consistent with the other experiments if a realistic
DM halo, with a mild dark disc component and platykurtic velocity distributions, as suggested by recent cosmological
simulations with baryons, is considered.

\subsubsection{Inelastic scenario}

For the strong dark disc halo model as for the standard Maxwellian halo, exclusion limits from CDMS
and CRESST leave us with the same two solutions to the DAMA puzzle (see Fig.~\ref{fig:DAMA2}).
Indeed, for $\delta > 30$~keV, the minimum velocity $v_{min}$ for a recoil is above 200~km/s,   
so that the low velocity enhancement for the strong dark disc does not contribute to the signal.

Things dramatically change when more realistic distributions are considered, either a slowly co-rotating
Tsallis halo or the model with a mild Tsallis dark disc.
While the low mass solution, $M_{DM} < 100$~GeV completely disappears, the high mass solution with
$M_{DM} \simeq few$~TeV is strongly enhanced.
Again, the improvement in compatibility between DAMA and other null experiments is mainly due
to the sharper drop of the high velocity tail in Tsallis distributions.
However, we notice that the mild dark disc setup gives a solution with a significantly smaller mass,
compared to the globally co-rotating Tsallis halo.

\subsection{Indirect Detection}

\begin{figure*}[t]
\begin{center}
\begin{tabular}{ccc}
\includegraphics[width=\columnwidth, clip]{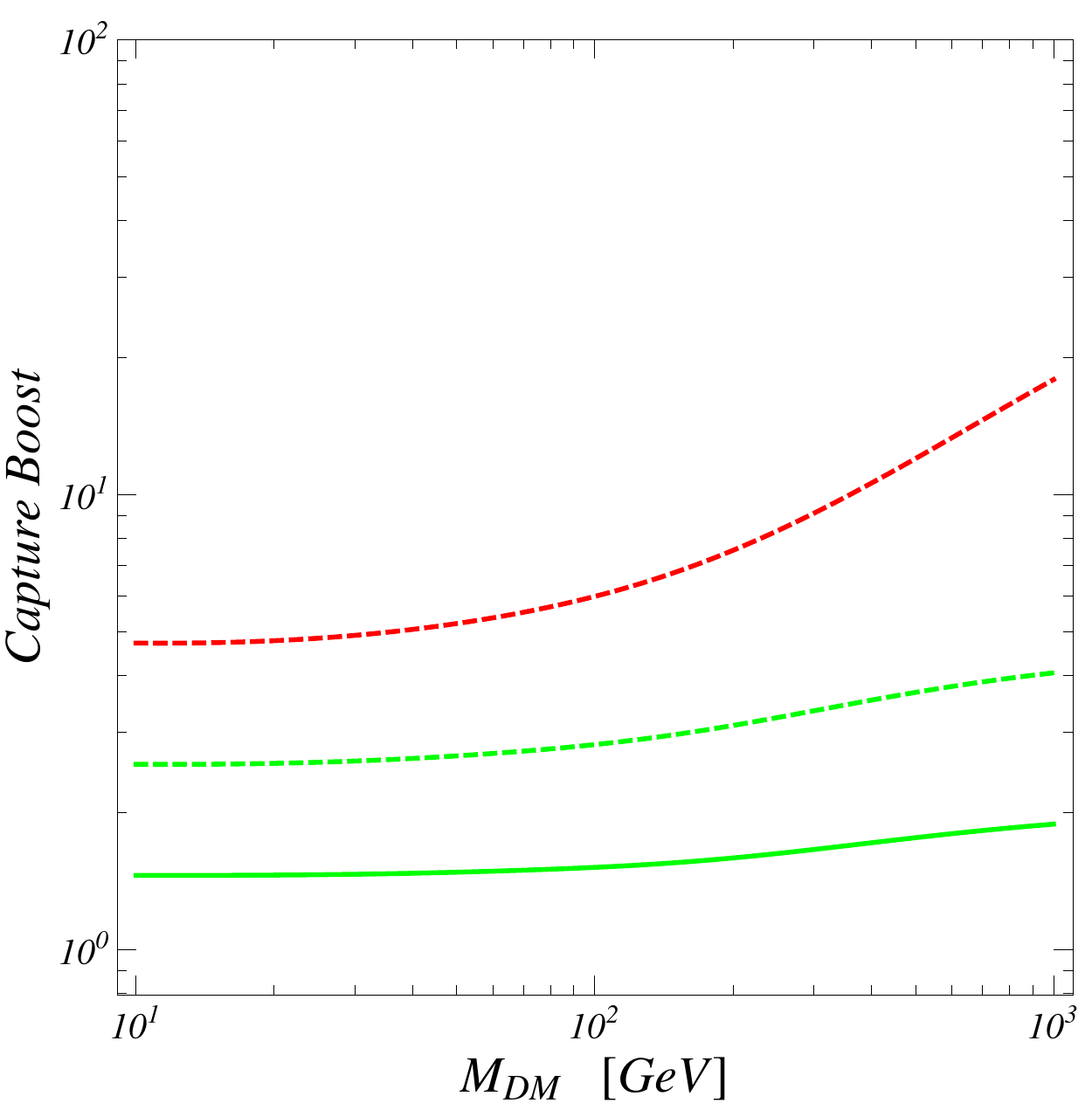}& ~~ &
\includegraphics[width=\columnwidth, clip]{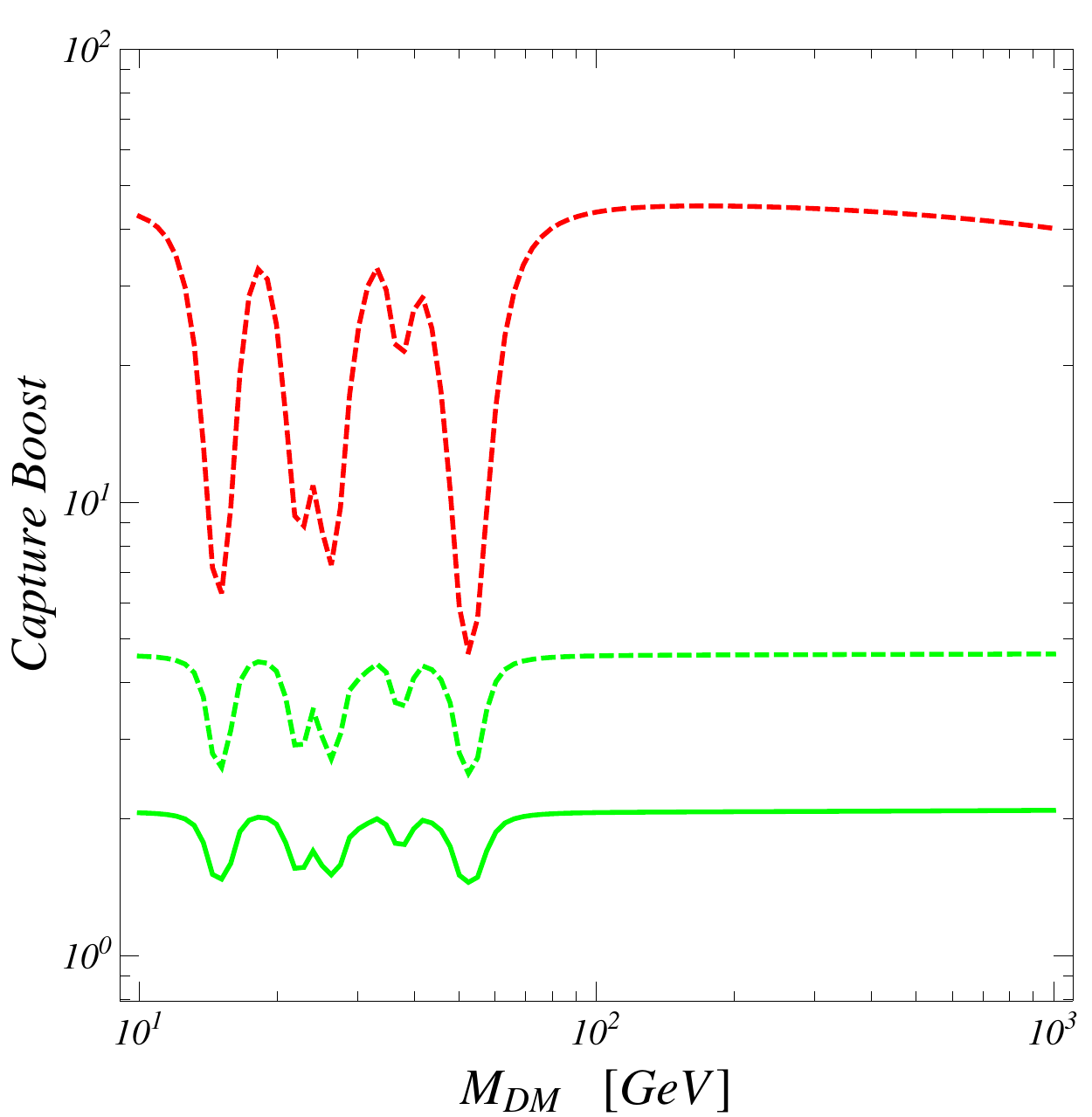}\\
\end{tabular}
\caption{\small \it 
{\rm \underline{Indirect detection}} :  Enhancement factor of the DM capture rate in the Sun (left) and in the Earth (right) compared to the
prediction for a standard Maxwellian halo, as a function of the DM mass.
\emph{Red dashed} : halo with a strong dark disc,
\emph{Green solid} : halo with a mild dark disc and $\rho_D/\rho_H = 1/3$,
\emph{Green dashed} : halo with a mild dark disc and $\rho_D/\rho_H = 1/1$.
}
\label{fig:CB}
\end{center}
\end{figure*}

Observation of very high energy neutrinos from the Sun or from the center of the Earth would be a convincing proof
of the existence of annihilating DM.
The flux of these neutrinos is dictated by the capture rate $C$ of the celestial body, rather than by the annihilation cross-section.
Indeed, the number of DM particles inside the body obeys the equation~\cite{Jungman:1995df}
\be
\frac{dN}{dt}=C-C_A N^2 \quad ,
\ee
where $\Gamma_A \equiv C_A N^2/2$ is the annihilation rate. Its solution is given by 
\be
\Gamma_A = \frac{C}{2} \tanh^2 (t/\tau) \quad ,
\ee
for constant $C$ and $C_A$, with the time constant $\tau = (C C_A)^{-1/2}$. 
Typically, $\tau \sim 10^8$~yr for the Sun and $\tau \sim 10^{11}$~yr for the Earth.
The age of the solar system is around 4.5 billion years, therefore $t/\tau \gg 1$ and $\Gamma_A = C/2$ for the Sun
while $t/\tau \ll 1$ and $\Gamma_A \propto C^2$ for the Earth. 

The capture rate per unit shell volume by an element with mass $M_N$, number density $n$ and zero-momentum cross-section $\sigma_0$ with the DM, 
is given by~\cite{Gould:1987ir,Jungman:1995df}
\bea
\frac{dC}{dV} = \frac{\rho_{DM}}{M_{DM}} \int_0^{v_{max}} dv_\infty \frac{{\rm f_W}(v_\infty)}{v_\infty} \; v \; \Omega^-_{v_{esc}}(v) \quad , \\
v \Omega^-_{v_{esc}}(v) = \sigma_0 n \frac{M_N}{2\mu^2} 
\int_{Q_{min}}^{Q_{max}} F^2(Q)dQ \quad ,
\eea
where ${\rm f_W}$ is the normalized DM velocity distribution in the Sun-frame, at `infinity' (\ie neglecting the distortions induced
by the gravitational potential of the celestial body), $v$ and $v_{esc}$ are a DM velocity and the value of escape velocity 
for the given shell, they are related to the velocity at infinity by $v^2=v_{esc}^2+v_\infty^2$,
$v_{max}=\sqrt{\beta_-}v_{esc}$ is the maximum velocity for which a capture is possible, 
$\mu$ is the WIMP-nucleus reduced mass, $Q_{min}=M_{DM} v_\infty^2/2$ and $Q_{max}=\beta_+ M_{DM} v^2/2$
are minimal and maximal values of the recoil energy for capture to be effective,
$F(Q)$ is a nuclear form factor such that $F(0)=1$,
and $\beta_\pm=4M_{DM}M_N/(M_{DM} \pm M_N)^2$.

To calculate the capture rate in the Sun, we use the density profile extracted from the reference standard solar model of Ref.~\cite{Basu:1999aa}, a chemical composition as found in Ref.~\cite{Jungman:1995df}, and an exponential form factor which largely reduces capture on heavy elements like iron. The escape velocity for the Sun ranges
from $v_{esc}^{max}=1395$~km/s in the center to $v_{esc}^{min}=618$~km/s at the surface. For heavy WIMPs, the impact of the escape velocity variation on the capture rate is very significant.
To calculate the capture rate in the Earth, we take a two layers (core + mantle) approximation model for the internal structure, with a different chemical composition for each layer~\cite{Allegre1995515}. 
The escape velocity for the Earth is considerably smaller ($v_{esc}^{max}=14.8$~km/s and $v_{esc}^{min}=11.2$~km/s), so that form factors can be safely neglected.
Here we consider only scalar interactions. 
Our purpose is to compare the predictions between different halos, rather than get a precise evaluation of the signal for a given halo.
Therefore, we calculate the capture enhancement for models with strong and mild dark discs compared to the capture rate for the standard
Maxwellian halo (Fig.~\ref{fig:CB}). In this way, we avoid many difficulties that plague the precise calculation of the neutrino flux,
in particular that coming from the Earth~\cite{Bruch:2009rp}.

For the Sun, it appears that the capture enhancement is moderate, even with a strong dark disc. 
Our result seems to be in conflict with the the Fig.~1 of Ref.~\cite{Bruch:2009rp}, 
where enhancements factors larger than 10 occur for masses smaller than 100~GeV.
The authors agree on the fact that the enhancement that can be read off their Fig.~1,
which is for the muon flux at the Earth's surface, should originate from a similar boost value
in the capture rate, as the annihilation rate of dark matter has reached a steady equilibrium
inside the Sun. Their calculation is entirely numerical, and details about form factors, or cross-sections 
are not given, which makes the comparison difficult.

In this work, we have focused on the capture boost for scalar interactions.
For a dark matter mass around 10~GeV, it appears that the capture boost $CB$ is well approximated by the
following zeroth order analytical limit
\be
CB \simeq \frac{\rho}{\rho_{SH}} \frac{I({\rm f_{W}})}{I({\rm f_{W,SH}})} \quad ,
\label{CB}
\ee
where $\rho$ and ${\rm f_W}$ are the local density and normalized velocity distribution 
for the dark matter halo considered, whereas the same quantities with a subscript SH refer
to the Standard Maxwellian halo, and
\be
I({\rm f_{W}}) = \int_0^\infty \frac{{\rm f_W}(v_\infty)}{v_\infty} \, dv_\infty 
\ee
Indeed, for $M_{DM} \simeq 10$~GeV, it can be checked that the capture rate is dominated 
by collisions on Helium, Oxygen and Iron, these are characterized by a maximum velocity $v_{max}$
larger than 700~km/s, and a form factor reduction smaller than 25\%.
The values of the capture boost as calculated with Eq.~(\ref{CB}) are 4.47 for the strong
dark disc halo model, and 2.49 and 1.44 for the mild dark disc halo models with $\rho_D/\rho_H=1$~and~$1/3$ 
respectively, in very good agreement with the exact values in Fig.~\ref{fig:CB}.
The capture boost becomes significantly larger than this approximation only for dark matter masses 
much smaller than the hydrogen atomic mass, or much larger than the iron atomic mass,
as the maximum velocity for a capture $v_{max}$ decreases to values where the 
low velocity strong enhancement due to the dark disc comes into play.

For the Earth, the capture enhancement is larger for the strong dark disc, up to a factor 50, and shows dips located around the resonances
with nuclei that are present with a large abundance, mainly Oxygen, Magnesium, Silicon, Iron and Nickel.
Finally, we note that for both the Sun and the Earth, the enhancement factor is limited below around two for the mild dark disc halo.
Therefore, if future N-body simulations confirm that realistic DM halos for the Milky Way have similar characteristics than our mild dark disc
model, then no spectacular boost is to be expected in neutrino indirect searches from the Sun and the Earth.

\section{Conclusions \& Perspectives}

In this letter, we have investigated the possible characteristics of the dark disc, in connection with results from N-body simulations with baryons.
We argue that its contribution to the local density is mild, and that its velocity dispersion is quite large.
Consequences for direct and indirect detection are studied. Signal enhancements are limited, but some subtle effects lead to an 
improvement of compatibility between DAMA and other null experiments. 

Both direct detection experiments and neutrino telescopes probe the local DM environment, they are also both sensitive to the DM elastic scattering
cross-section. However, a modification of the DM phase-space distribution affects differently these signals.
If the DAMA modulation signal is taken seriously, both the DM mass and its spin-independent cross-section would be more or less determined.
This analysis shows how neutrino telescopes provide complementary information, the indirect signal would further give indications on the local halo structure. 

\section*{Acknowledgments}

This work is funded in part by IISN and by Belgian Science Policy (IAP VI/11).

\bibliographystyle{unsrt}
\bibliography{DD}

\end{document}